\documentclass[fleqn,usenatbib]{mnras}

\usepackage{newtxtext,newtxmath}
\usepackage{booktabs}
\usepackage{float}
\usepackage{graphicx}
\usepackage{xcolor}

\usepackage[T1]{fontenc}

\DeclareRobustCommand{\VAN}[3]{#2}
\let\VANthebibliography\thebibliography
\def\thebibliography{\DeclareRobustCommand{\VAN}[3]{##3}\VANthebibliography}

\usepackage{graphicx}	
\usepackage{amsmath}	

\newcommand{\Msun}{\mathrm{M}_{\odot}}

\title[]{Star formation rate density as a function of galaxy mass at $z <  0.2$ with MUSE and GAMA surveys}

\author[G. G. Murrell and I. K. Baldry]{
Gregory G. Murrell,$^{1}$\thanks{E-mail: G.G.Murrell@2021.ljmu.ac.uk}
I. K. Baldry,$^{1}$
\\
$^{1}$Astrophysics Research Institute, Liverpool John Moores University, 146 Brownlow Hill, Liverpool L3 5RF, UK\\
}

\date{submitted October 2024, revised March 2025}

\pubyear{2025}

\begin{document}
\label{firstpage}
\pagerange{\pageref{firstpage}--\pageref{lastpage}}
\maketitle

\begin{abstract}
The star formation rate density (SFRD) is an important tool in galaxy evolution that allows us to identify at which cosmic time galaxies are more efficient at forming stars. For low-mass star-forming galaxies, the SFRD as a function of stellar mass can be straightforwardly related to the galaxy stellar mass function (GSMF). Given the uncertainty of the GSMF at the low-mass end, due to the challenges in observing dwarf galaxies, deriving the SFRD with respect to mass may be crucial to understand galaxy formation. 
Measurement of SFRD is more complete than number density in a cosmological volume because galaxies with higher SFR are easier to detect and characterize. 
In this work, the SFRD is derived using two different samples, one using the MUSE WIDE and MUSE Hubble Ultra Deep Field IFU spectroscopic surveys, and another using the GAMA spectroscopic survey. The first sample comprised a total of 27 star-forming galaxies at $z < 0.2$ (H$\alpha$ selected), whereas the second contained 7579 galaxies at $z < 0.06$ ($r$-band selected). The star formation rates are derived from measurements of the H$\alpha$ emission line fluxes for the first sample, and using MagPhys SED fitting for the second one.  The results show the behaviour of the SFRD to the lowest stellar masses of $10^{5.5} \Msun$, consistent with a constant slope (in log SFRD versus log stellar mass) and thus no turn-over in the GSMF. 
\end{abstract}




 \section{Introduction}

The star formation rate density (SFRD) is defined as the total star-formation rate (SFR) averaged over a given comoving volume of the universe observed at a given redshift \citep[]{madau1996high}. Understanding the evolution of the SFRD across cosmic history may unravel crucial insights into how galaxies grow and evolve, constraining cosmological models \citep[]{schaye2010physics}. When measuring its contribution as a function of redshift, it allows us to understand at which cosmic time, star formation is more favourable in galaxies; whereas if analysed with respect to galaxy stellar mass, the SFRD may provide useful information on the role of galaxy mass in star formation \cite[]{gilbank2010local}. 

SFRD measurements over cosmic time have become a common practice in modern astrophysics in the past three decades. Numerous studies have estimated the shape of SFRD, with literature agreeing that it peaks at around $z \sim $ 2-3, declines in more recent epochs, with another sharper drop at $z > 8$  \citep[]{bouwens2015uv}. \cite{lilly1996canada} pioneered the analysis of SFRD, conducting the first systematic work to combine a large and deep spectroscopic redshift survey with multi-wavelength photometry, and deriving the luminosity function and SFRD to redshifts of $z \sim 1$. \cite{hopkins2006normalization} showed how the cosmic star formation history may be disclosed by measurements of the SFRD, as it shows in which epochs star formation is more favourable. HST data was used by \cite{bouwens2015uv} to push the limits of SFRD to the highest redshifts, as they derived the SFRD from UV luminosity functions to redshifts of $z \sim 10$, showing a sharp decrease in star formation at redshifts of $z > 8$. \cite{madau2014cosmic} reviewed over 200 SFRD measurements from the literature, reporting methodologies and results, alongside connecting the observational results with cosmological theories.

Although there has been a vast number of studies on how SFRD evolves over time, the same cannot be said on how SFRD relates to other intrinsic galaxy properties. Understanding the relation between SFRD and other physical properties may give us additional information on how galaxies evolve and assemble \citep[]{madau2014cosmic}. For example, \cite{james2008h} have shown that SFRD varies significantly with galaxy morphology, highlighting the importance of galaxy type in shaping star formation activity. 
\cite{schulze2021palomar} measured CCSNe rates (closely related to SFRD) as a function of galaxy mass for different types of supernovae. 
If analysed with respect to mass distribution, the SFRD provides useful insights of how galaxies of given masses contribute to the total star-formation-rate budget across cosmic time, revealing mass-dependent trends in star formation history \citep{gilbank2010redshift,khostovan24}. Studies such as \cite{drake2015evolution} have further examined this relation, showing how star formation evolves with stellar mass between redshifts $z = 1.46$ and $z = 0.63$. 

For low-mass star-forming galaxies, assuming that the mean specific star formation rate is approximately constant for masses smaller than $\sim 10^9 \Msun$, measurements of the SFRD can give an estimate of the galaxy stellar mass function (GSMF) \cite[]{sedgwick2019galaxy}. This assumption is confirmed by the analysis of H$\alpha$ emission as a function of galaxy stellar mass \cite[]{james2008halpha} and by unquenched low-mass galaxies, dwarf galaxies whose cold gas reservoir has not been emptied by any physical mechanism, as they have been forming stars at a quasi-constant rate, meaning that it is reasonable to expect that their \textit{mean} specific SFR is $\sim 10^{-10} yr^{-1}$ \cite[]{van2001evolutionary}.

\citet{gilbank2010local} calculated the SFRD using H$\alpha$, [OII], and $u$-band luminosities from the Sloan Digital Sky Survey (SDSS) \cite[]{york2000sloan} coupled with UV data from the Galaxy Evolution EXplorer (GALEX) satellite \cite[]{budavari2009galex}, deriving the SFRD with respect to stellar mass down to masses of $\sim 10^8 \Msun$. However, the derivation of the SFRD at lower masses has proven to be a challenge due to the extremely low surface brightness of dwarf galaxies \cite[]{cross2002bivariate}. \cite{sedgwick2019galaxy} introduced a novel approach, using core-collapse supernovae (CCSNe) as "signposts" to low surface brightness galaxies to constrain their abundance and contribution to the total star formation rate density (SFRD) at lower masses. Rates are converted to SFRDs using the expected CCSNe per unit mass of star formation. The work from \cite{sedgwick2019galaxy} suggests that a greater fraction of star formation occurs in low-mass, low-surface brightness galaxies than previously thought. Nevertheless, more sensitive surveys are needed to fully capture the star formation activity and extend SFRD measurements to smaller and fainter systems.

Previous studies of SFRD have used used emission lines and broadband photometric tracers to extend SFRD measurements to low stellar masses across cosmic time, with the H$\alpha$ emission line being one of the best indicators of star formation in a galaxy, since young, massive stars with ages $<$ 20 Myr produce copious amounts of ionising photons that ionise the surrounding gas \cite[]{davies2016gama}. This is a great advantage when calculating the SFR of a galaxy, as only young stars are taken into account and thus it is largely independent of SF history \cite[]{kennicutt1998global}. Additionally, compared to other emission lines, H$\alpha$ is the most directly proportional to the ionizing UV stellar spectra at $\lambda$ $<$ 912 \AA\ and it is not as susceptible to the metal fraction present in the gas \cite[]{tresse1998halpha}. A limitation of using H$\alpha$ as a star-formation rate indicator is that the aperture based spectroscopy only probes the central regions of nearby galaxies \cite[]{davies2016gama}, but this can be corrected with the appropriate calibration or with integral-field unit (IFU) spectroscopy.

As modern IFU surveys can provide complete spatial and spectral coverage, detecting emission lines even for the faintest and most diffuse sources that lack continuum, they can be used to probe the faintest galaxy populations, extending SFRD measurements to the lowest masses \citep[]{bacon2015muse, herenz2017muse}. The MUSE Hubble Ultra Deep Field (HUDF) \cite[]{bacon2023muse} and MUSE Wide \cite[]{urrutia2019muse} are optical IFU spectroscopic surveys from the MUSE collaboration that can reach unprecedented depths compared to similar spectroscopic surveys.

In this work, the low-mass galaxy population at redshift $\leq$ 0.2 is characterised using MUSE Wide and MUSE HUDF, deriving their SFRs through the analysis of their H$\alpha$ emission line fluxes. Using this information, the SFRD with respect to stellar mass of the population for the combined MUSE sample is calculated. The SFRD is also derived using the Galaxy and Mass Assembly (GAMA) survey. 
{GAMA is a much larger spectroscopic survey but is selected using broadband SDSS photometry and is 
expected to be significantly incomplete to low-surface-brightness galaxies below about $10^8 \Msun$ \citep{baldry2012galaxy}. 
Results of both approaches are presented to determine the contribution of the low-mass galaxy population to the cosmic star formation rate density.}

A description of the data sets used is presented in \S~\ref{sec:data}, 
the principles of the SFRD measurements are presented in \S~\ref{sec:sfrd-principles},
sample selection and processing in \S~\ref{sec:analysis},
and the main results in \S~\ref{sec:results}; 
with discussion and summary in the final sections. 
An $\Omega_m=0.3$ flat-$\Lambda$ cosmology with $H_0=70$ is assumed for distances and volumes.

\section{Data sample and surveys description}
\label{sec:data}

In this section, we describe the source data and the different surveys used in this study. Each subsection details the characteristics of the each survey, explaining the data release used, how the measurements were derived, and which catalogues are used in this study.
\subsection{MUSE Wide}
\label{sec:MUSE_Wide} 
The MUSE Wide Survey is a large spectroscopic survey that uses the Multi-Unit Spectroscopic Explorer (MUSE) instrument on the Very Large Telescope (VLT) to observe galaxies in the distant universe \citep{bacon2010muse}. The MUSE Wide survey is a blind, 3D spectroscopic survey that provides, like many other extragalactic surveys, a “wedding-cake” approach, and is expected to cover 100 × 1 arcmin\textsuperscript{2} MUSE fields by the end of the survey \cite[]{urrutia2019muse}. MUSE-Wide mainly covers parts of the CDFS and COSMOS regions that were previously mapped by HST in several bands to intermediate depths, by GOODS-South in the optical \cite[]{giavalisco2004great} and by CANDELS in the near infrared \citep{grogin2011candels,koekemoer11}. The survey is designed to cover the whole field of view continuously, so that it is not restricted to a photometric pre-selection for identification and classification of objects in the sky \cite[]{bacon2017muse}.

The first data release (DR1) consists of 44 CANDELS-CDFS fields, with observations carried out in nominal mode (each spectrum spans from 4750 – 9350 Å in wavelength range), and 0.2''×0.2'' spatial and 1.25 Å wavelength sampling, which is the default for MUSE \cite[]{urrutia2019muse}. The survey used a combination of spectroscopic and photometric measurements to derive the stellar masses of the galaxies in their catalogue. The photometry was taken from \cite{skelton20143d}, whilst the software FAST (Fitting and Assessment of Synthetic Templates, \citealt{kriek2018fast}) was used for the spectral energy distribution (SED) fitting model. FAST determines the best fit parameters using $\chi^2$ minimization from a set of model SEDs and an analysis grid describing several stellar population models.

The catalogues used in this study were retrieved from the MUSE Wide DR1 database, which is available on their website (\href{https://musewide.aip.de/metadata/musewide_dr1/}{MUSEWIDE\_DR1}). The two catalogues used in this work MUSE, one being the MUSE WIDE main table (musewide\_dr1.mw\_44fields\_main\_table), from which the columns containing galaxy properties such as redshift and stellar mass, and their associated source with the \cite{skelton20143d} catalogue, were extracted (columns names respectively: \textsc{z, stellar\_mass, skelton\_id, skelton\_sep}). The other catalogue used is the MUSE WIDE emission line table (musewide\_dr1.mw\_44fields\_emline\_table), containing information about line fluxes at different Kron radius apertures (column used \textsc{f\_2kron}, which comprises values for emission line fluxes extracted in an aperture equivalent to Kron radius $\times$ 2). The emission line and source identification was performed using \textsc{LSDCat} \citep{herenz2017lsdcat}, an automated detection package for emission lines in wide-field integral-field spectroscopic datacubes. The full description of the data release and catalogues is reported in the documentation from \cite{urrutia2019muse} and \cite{herenz2017lsdcat}.

\subsection{MUSE Hubble Ultra Deep Field}
\label{sec:MUSE_HUDF}

The MUSE Hubble Ultra-Deep Field (HUDF) survey is a deep spectroscopic survey of the Hubble Ultra Deep Field region. The second data release (DR2) provides the deepest IFU spectroscopic survey to date, alongside excellent 3D content, wide spectral range, and outstanding spatial and spectral resolution \cite[]{bacon2023muse}. The survey is based on three MUSE datasets at various depths: MOSAIC (3x3 arcmin\textsuperscript{2}, 10 hours), UDF-10 (1 arcmin\textsuperscript{2}, 31 hours) and MXDF (1 arcmin diameter, 141 hours). The stellar masses of the galaxy sample of the survey were derived using SED fitting. In order to perform the SED fit with sufficient constraints, photometry from the HST R15 \citep[]{rafelski2015uvudf} catalogue was used, as the catalogue contains 11 photometric bands ranging from the NUV (0.21µm) to the WFC3/IR (1.5µm). The stellar masses were calculated using two different SED fitting codes: the high-z extension of MagPhys \citep{da2008simple,da2015alma}, with minimum stellar mass of $10^6 \Msun$, and Prospector \citep[]{johnson2021stellar}. However, it is reported that Prospector tends to derive higher values for stellar masses compared to MagPhys, with a median offset of 0.25 dex, which is a known characteristic of Prospector \citep{leja2020new}. 

From the  MUSE HUDF survey, the data from the DR2 main table (dr2\_main\_09) is used for the analysis. The table contains galaxy properties like redshift and stellar mass derived using MagPhys, which can be accessed via their respective columns 'Z' and 'MASS\_MAG'. For each galaxy, the table also contains values for emission line fluxes, with the H$\alpha$ flux values used being stored in the 'HALPHA\_EMI\_FLUX' column. For the MUSE HUDF survey, the source detection and classification was performed using the blind detection software ORIGIN, capable of detecting faint line emitters in MUSE datacubes \citep{mary2020origin}. The dataset is publicly available and may be retrieved from the collaboration website (\href{https://amused.univ-lyon1.fr/project/UDF/HUDF/}{MUSEHUDF\_DR2}). For full description of the survey and catalogues, please refer to \cite{bacon2023muse}.

\subsection{GAMA}
\label{sec:GAMA}
The Galaxy And Mass Assembly (GAMA) survey \citep{liske2015,baldry2018galaxy,driver2022galaxy} is a spectroscopic redshift and multi-wavelength photometric survey designed to study galaxy evolution. The survey is divided into five regions: three equatorial regions (G09, G12, G15) each of 60.0 deg\textsuperscript{2} and main survey limit of r\textsubscript{AB} $<$ 19.8 mag, and two southern regions of 55.7 deg\textsuperscript{2} (G02) and 50.6 deg\textsuperscript{2} (G23). 
Here, we use the equatorial regions, which cover an area of 179 deg\textsuperscript{2} after accounting for masking around bright stars. 

The fourth data release of GAMA (DR4) is the final release \citep{driver2022galaxy}. Compared to MUSE Wide and MUSE HUDF, GAMA is shallower. Moreover, it relies on fiber spectroscopy, which may result in an overestimation of emission line fluxes and SFRs for small or low surface brightness galaxies when applying aperture corrections \citep{richards2016sami}. However, GAMA benefits from a much wider area and larger catalogue, comprising 21-band photometric data with spectroscopic line emission measurements for star formation indicators OII and H$\alpha$, containing information for stellar mass estimates of 198223 galaxies, and measured redshifts for 196402.

The survey data is organised into data management units (DMUs). The DMUs used in this work are: 
SpecCat, containing the spectra and redshifts from all the curated spectroscopic data of the GAMA survey, with the primary choice of redshift extracted using the automatic code AUTOZ \citep{baldry2014galaxy}; 
StellarMasses, which provides measurements of the total stellar mass of the sources \citep[]{taylor2011galaxy}; 
MagPhys \citep{da2008simple}, that contains estimates of a number of key parameters including stellar mass, dust mass, and star-formation rate (SFR) \citep[]{driver2018gama}. 
For the descriptions of other DMUs, please refer to \cite{baldry2018galaxy} and \cite{driver2022galaxy}.
{Note that we prefer the MagPhys estimates of the SFRs compared to the H$\alpha$ fluxes \citep{hopkins2013,gordon2016galaxy} because, in the latter case,
the fluxes are indirectly derived from equivalent widths and $r$-band photometry which can result in unverified high-SFR outliers \citep{davies2016gama} that significantly impact SFRD measurements.}

\section{Derivation of the Star Formation Rate Density with the $1/V_{max}$ method}
\label{sec:sfrd-principles}


\subsection{Star Formation Rate from H$\alpha$ emission lines}
\label{sec:ha_line}
Star forming galaxies produce very specific emission lines in their spectra that can be easily characterised. H$\alpha$ emission lines are one of the best indicators for star forming galaxies, since only young, massive stars with ages $<$ 20 Myr produce vast amounts of photons that ionise the surrounding gas \citep{davies2016gama}. This is a great advantage when calculating the SFR of a galaxy, as only young stars are taken into account and thus it is largely independent of SF history \citep{kennicutt1998global}. Assuming a Kroupa initial mass function \citep[]{kroupa2001variation}, the relation between the H$\alpha$ luminosity and star formation rate may be obtained with the following calibration \citep{calzetti2013star}: 
\begin{equation}
    \textrm{SFR$_{H\alpha}$} = 5.5 \times 10^{-42} L(H\alpha)
    \label{eq:SFRCalzetti}
\end{equation} with SFR$(H\alpha)$ in M\textsubscript{\(\odot\)} yr\textsuperscript{-1} and $L(H\alpha)$ in erg s\textsuperscript{-1}. The calibration constant is valid provided that star formation has remained constant over timescales $>$ 6 Myr, with no dependency on long timescales \citep{calzetti2013star}. This calibration does not take into account AGN contamination, which could theoretically result in an overestimation of SFR(H$\alpha$). However, for low-mass galaxies AGN contamination is not a significant issue \cite[]{kauffmann2003host}

\subsection{Star Formation Rate Density with the $1/V_{max}$ method}
\label{sec:sfr_vmax}
{The SFRD with respect to mass can be calculated using the $1/V_{max}$ method. The $1/V_{max}$  method \citep{felten1976schmidt} corrects for survey biases by weighting galaxies according to the maximum comoving volume in which they could be detected. This ensures that fainter galaxies, observable over smaller volumes, contribute proportionately more to the overall density estimate. The comoving volume may be derived using the following equation:}
\begin{equation}
    V_{max} = \frac{\Omega}{3} \left( D^{3}_{max} - D^{3}_{min} \right)
    \label{eq:volume}
\end{equation} 
where $\Omega$ is the solid angle of the FOV of the survey, and $D^{3}_{max}$ and $D^{3}_{min}$ are the maximum and minimum comoving distance for which the object could be observed given its luminosity and considering the survey limits. 
{Note for a redshift-limited sample, 
the brighter sources that are limited only by the redshift range will have the same $V_{max}$.}

If more than one survey is used, the $V_{max}$ estimate has to consider the total volume of the surveys. This can be done assuming that a source may be observed at any given point of the surveys. Therefore, if two surveys are used, the total $V_{max}$ may be calculated by adding the $V_{max}$ of a galaxy in each survey: 
\begin{equation}
    V_{tot} = V_{max, 1} \, + \, V_{max, 2}
    \label{eq:volume_tot}
\end{equation} 
with $V_{max, 1}$ being the comoving volume of the source in its "parent" survey, and $V_{max, 2}$ the comoving volume if the galaxy had been observed in the other survey. 

Once the total comoving volume has been defined, the $1/V_{max}$ method may be used to derive the SFRD as a function of mass in the $m^{th}$ bin using the following relation: 
\begin{equation}
    \rho_{SFR,m} = \frac{1}{\Delta \log M} \sum \frac{SFR_i}{V_{max,i}}
    \label{eq:sfrd}
\end{equation} where for the $i^{th}$ galaxy, $SFR_i$ is its SFR, $V_{max, i}$ is its maximum volume from which the object could still be detected, and $\Delta \log M$ is the bin width.
In other words, this provides a binned estimate of 
$d \mathrm{SFRD} / d \log M$ 
in units of solar masses per year per Mpc$^3$ per dex. 

\subsection{Estimation of $D_{max}$}

In order to correctly apply the 1/V$_{max}$ method described in Sec.~\ref{sec:sfr_vmax}, it is necessary to estimate the maximum distance at which a source can be observed given its luminosity. This can be done in different ways. For the sample from MUSE WIDE and MUSE HUDF, D$_{max}$ was derived by defining a minimum flux detection limit (see Sec. \ref{sec:det_lim} for derivation). If the detection limit is defined as the minimum flux that a source must have in order to be included in the sample ($F_{lim}$), one can work out the maximum luminosity distance ($D_{L,max}$) at which it can be observed by rearranging the relationship between bolometric luminosity and flux: 
\begin{equation}
    F_{lim} = \frac{L}{4\pi D_{L,max}^2} \: .
\end{equation}
In the case of bolometric flux, i.e.\ with no k-correction, then 
\begin{equation}
    D_{L,max} = D_{L,obs} \sqrt{\frac{F_{obs}}{F_{lim}}} \: .
\end{equation}
where $F_{obs}$ is the observed flux, and $D_{L,obs}$ is the observed luminosity distance. 
This applies to line emission. 

The $D_{max}$ values for the galaxies in the GAMA survey were derived using redshift and magnitude data from SDSS. The procedure requires calculating k-corrections, at a range of redshifts, in order to iterate toward an estimate for the maximum distance. 
The maximum distance can be obtained from the distance modulus given by:
\begin{equation}
    \textrm{DM}_{max} = \textrm{DM}_{obs} + m_{lim} - m_{obs} + dK
\end{equation} 
where DM$_{obs}$ is the distance modulus at the observed redshift, 
$m_{lim}$ is the magnitude limit of the survey, 
$m_{obs}$ is the observed magnitude of the source, and 
$dK$ is the differential K-correction.
The latter is given by 
\begin{equation}
    dK = k_{max,obs} + 2.5 \log (1 + z_{max})
\end{equation} 
where $k_{max,obs}$ is the k-correction of the observed galaxy for the effective band at $z_{max}$ (see \citealt{blanton03kcorr} for this nomenclature). This is obtained by iterating: starting from $dK = 0$ and determining a new $dK$ by interpolation from a vector of band-shifted k-correction values on each iteration. Note that $k_{obs,obs} = -2.5 \log (1 + z_{obs})$, ensuring that $dK$ converges to zero when $z_{max} = z_{obs}$. 

\subsection{Dust extinction}
\label{sec:dust_extinction}
In order to accurately measure the SFR, dust extinction corrections must be taken into account. H$\alpha$ line is situated in the optical spectrum, making it less sensitive to dust extinction compared to other star formation indicators such as UV light. However, even for local galaxies, dust effects on H$\alpha$ are significant \citep[]{koyama2015predicting}. A common practice to determine the levels of dust attenuation in galaxies is to measure the ratio between H$\alpha$ and H$\beta$, also known as the Balmer decrement. The theoretical value of the Balmer decrement, assuming that a optically thick star forming region for all Lyman lines greater than Ly$\alpha$, is equal to 2.86 \citep[]{calzetti2000dust}. Therefore, for a fixed electron temperature of $10^4$ K, one can assume that any deviation from the theoretical value is due to dust extinction \citep[]{groves2012balmer}.

The magnitude extinction $A_V$ from the observed Balmer decrement may be calculated using the relation described in \cite{osterbrock2006astrophysics}, with the E(B-V) colour excess that can be calculated directly from the Balmer decrement using the equation (see \citep{dominguez2013dust}): 
\begin{equation}
    E(B-V) = \frac{2.5}{k(\text{H}\beta) - k(\text{H}\alpha)} \times \log\left(\frac{(\text{H}\alpha/\text{H}\beta){\text{obs}}}{2.86}\right)
\end{equation} with the extinction coefficients $k(\text{H}\alpha)$ and $k(\text{H}\beta)$ coming from the reddening curve between H$\alpha$ and H$\beta$ described in \cite{cardelli1989relationship}. The dust correction factor is then derived with \cite{calzetti2000dust}: 
\begin{equation}
    \mathrm{Dust \: Correction \: Factor} =  10^{0.4 \: E(B-V) \: k(\mathrm{H}\alpha)}
\end{equation} Finally, the correction factor can be used to account dust extinction in the observed H$\alpha$ flux:
\begin{equation}
    H\alpha \: \mathrm{Flux_{Corr}} = H\alpha \:\mathrm{Flux_{Obs}} \times \mathrm{Dust \: Correction \: Factor}
\end{equation}

\section{Analysis}
\label{sec:analysis}

\subsection{Sample selection}

\begin{figure}
    \centering
    \includegraphics[width=8.5cm]{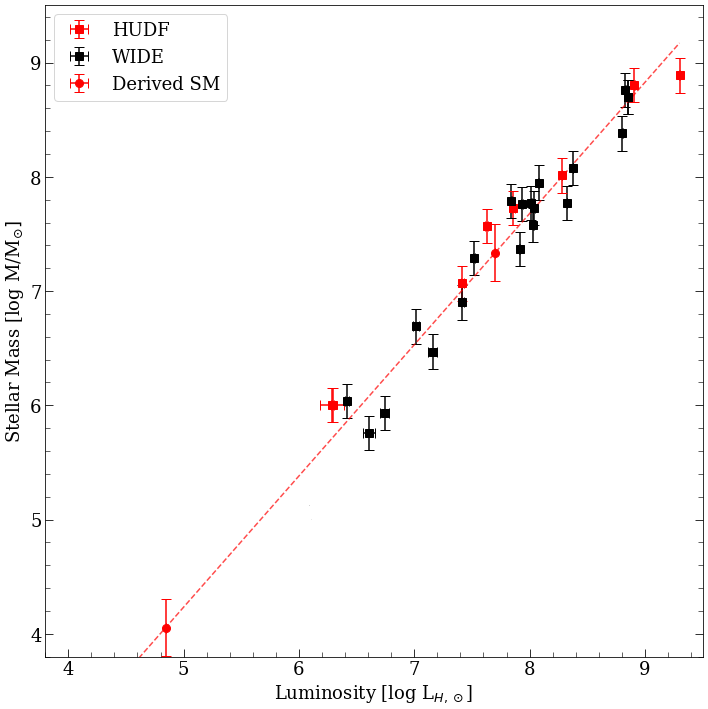}
    \caption{Mass-Luminosity relation of the galaxies in the sample. The luminosities are derived using \citet{van20143d} photometric catalogue. Using the log-log linear fit shown on this plot, the stellar masses of the sources without a valid mass on the catalogue were derived {(these two sources are shown with circles on top of the line). An indicative stellar mass
    uncertainty was used. This was obtained by calculating the RMS difference between two different estimates from the MagPhys and Prospector codes.}}
    \label{fig:ML_ratio}
\end{figure}

Initially, all the galaxies in the MUSE Wide and HUDF catalogues \citep{urrutia2019muse,bacon2023muse} that presented H$\alpha$ emission lines at redshift $\leq$ 0.2 were considered. Since the two areas covered by HUDF and WIDE overlay slightly, it was also ensured that there were no duplicated galaxies, resulting in a preliminary sample of 21 star-forming galaxies from MUSE Wide emission line catalogue and 10 from MUSE HUDF main catalogue. In order to verify the accuracy of the redshift measurements, AUTOZ, which is a fully automated redshift code that allows us to homogenise redshift measurements \citep{baldry2014galaxy}, was run over the entire area. The test resulted in no significant difference between the values reported in the MUSE Wide and HUDF catalogues, and the redshifts derived using AUTOZ.

MUSE Wide DR1 and MUSE HUDF DR2 used different photometry to derive the stellar masses values for the galaxies, the \cite{skelton20143d} and the HST R15 \cite{rafelski2015uvudf} catalogues respectively. The HST R15 \cite{rafelski2015uvudf} catalogue used in HUDF, does not contain values for F160W magnitude for all the sources in the catalogue. Therefore, photometry from the \cite{van20143d} catalogue was used instead, as it contained values for all the sources in the sample. In order to ensure consistency, the photometric catalogues the F160W magnitudes were compared. This allowed us to assume that using photometric values from either catalogue does not affect the derivation of the H-band luminosity of the sources.

Using the F160W magnitude values from the \cite{van20143d} catalogue, the H-band luminosities of all the galaxies in the sample were calculated. This information was then used to derive the stellar mass of two sources in the HUDF region that do not have a valid derivation for the stellar mass using MagPhys. In fact, Source ID6474 does not have a photometric value in R15, whereas the stellar mass of source ID8051 is likely extremely low. In order to include them in the sample, as they both present H$\alpha$ emission lines, their stellar masses were derived using the typical stellar mass-to-luminosity ratio of star forming galaxies. 

The relation between stellar mass and luminosity of the sample is shown in the plot in Fig. \ref{fig:ML_ratio}, showing a linear relationship in the log-log scale. Hence, the stellar mass of the two star forming galaxies was derived, with the assumption that they follow the stellar mass-luminosity relation.

\begin{figure}
    \centering
    \includegraphics[width=8.5cm]{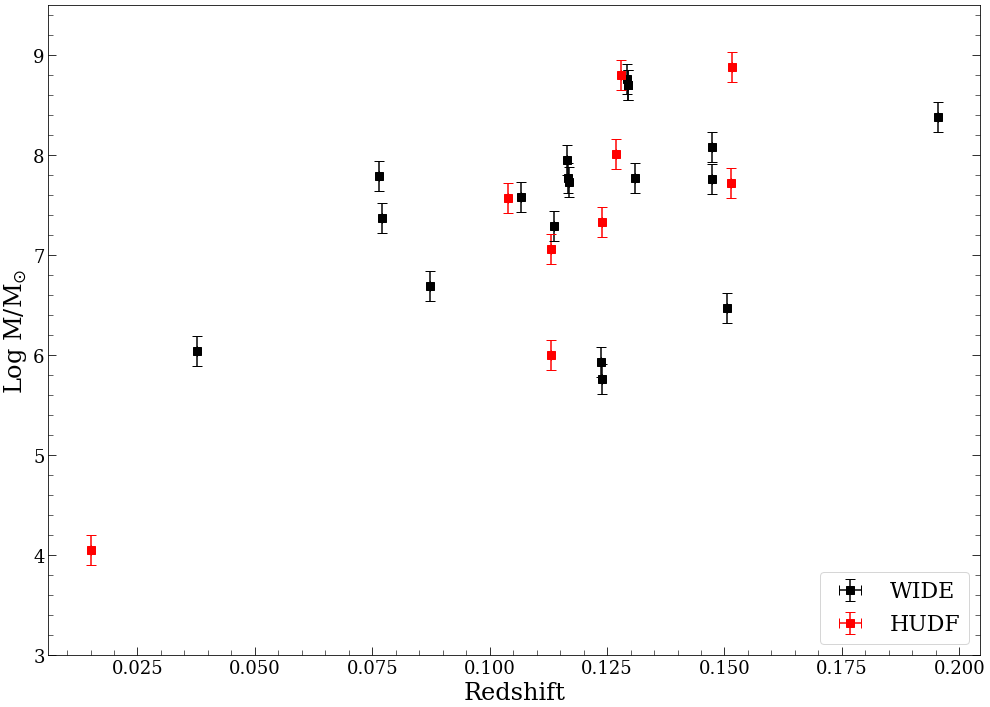}
    \caption{Stellar mass with respect to redshift for the combined sample of MUSE Wide emission line catalogue \citep{urrutia2019muse} and MUSE HUDF catalogue \citep{bacon2023muse}. From the plot, it can be seen that the sample is comprised by very low-mass star-forming galaxies, with all the galaxies with stellar masses below $\sim 10^9 \Msun$. Please note that the plot also shows ID8051 (red square, lowest stellar mass in the sample) as its mass was derived using the stellar mass-luminosity relation, but it is not used in the calculation of the SFRD as it falls just below the empirical detection limit.}
    \label{fig:mass_z}
\end{figure}

The stellar masses of the sample can be seen in Fig. \ref{fig:mass_z}, where the values of the masses with respect to redshift are shown. ID8051, whose stellar mass was derived using the mass-luminosity relation, appears to have an exceptionally low mass of $\sim 10^4 \Msun$, which, if confirmed to be accurate, would be one of the lowest stellar masses known of star-forming galaxies.

\subsection{Detection Limit}
\label{sec:det_lim}
Estimating the detection limits of the surveys is necessary in order to perform the $1/V_{max}$ method \citep{schmidt1968space}. In fact, because they affect the maximum volume within which objects can be detected, a failure to account for this bias can lead to incorrect estimates of the space density of the population being studied. However, the derivation of detection limits is often very complex, as there are many factors such as sky lines, atmospheric conditions, and aperture sizes that need to be considered. 

In order to simplify the derivation of the detection limit and the calculation of V$_{max}$, an empirical detection limit (i.e.\ sample-selection limit) was set by analysing all the emission line fluxes detected in the survey between 6563 and 7876 \AA, which is the region where H$\alpha$ is expected to be found at $z \leq 0.2$. A reasonable assumption would be that the highest number of observations will have values just above the detection limit, and that below the limit there will be a sharp decrease in detections, since the instrument would have more difficulty in detecting the signal below a certain threshold.

\begin{figure*}
    \centering
    \includegraphics[width=14cm]{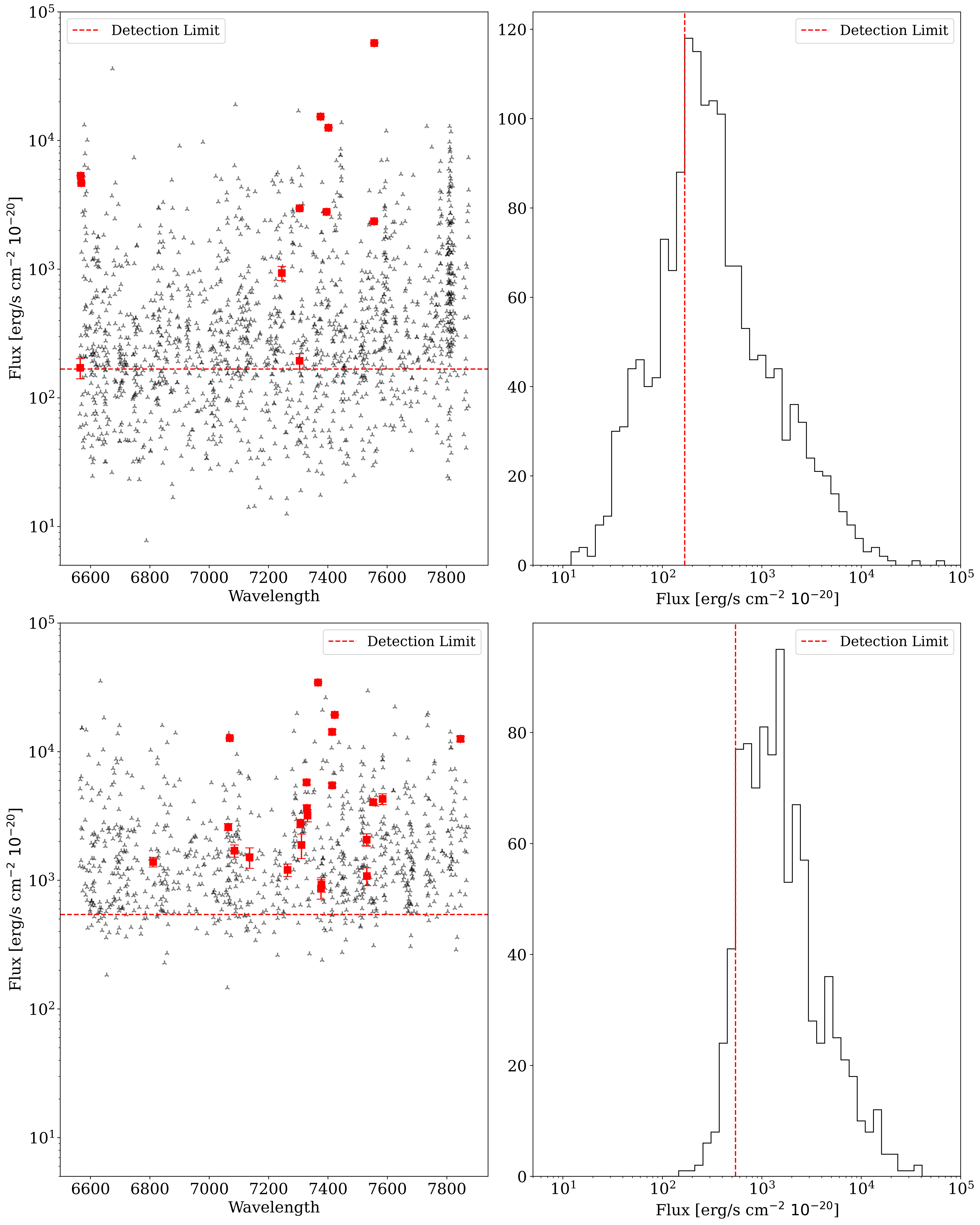}
    \caption{Derivation of the detection limit for the MUSE HUDF sample (top panels) and MUSE WIDE sample (bottom panels). On the left panels, it is possible to visualise all the emission lines (grey triangles) detected in the range of interest, with the red squares indicating the detected H$\alpha$ lines. It can be seen how MUSE HUDF is much deeper compared to MUSE Wide, as there are significantly more detections at lower fluxes. {On the right panels, the histograms of the line fluxes are shown. The red dashed lines indicate the sample selection limits.}}
    \label{fig:det_limit}
\end{figure*}

Using the histogram shown in Fig. \ref{fig:det_limit}, the detection limit was estimated. It can be seen that there is a significant decline in number of emission lines observed right before the peak in the histogram. The detection limits were thus set to be the cut-off values where the number of detections dropped, which can be seen in the plots as the red dashed line. The plots also show all the emission line fluxes detected between 6563 and 7876 \AA, with the red points representing the H$\alpha$ lines. The results show that all the H$\alpha$ emissions in MUSE Wide, and all but one in MUSE HUDF, are above the empirically estimated detection limits. The source below this limit (MUSE HUDF ID8061) is excluded from the estimation of the SFRD.

\subsection{Dust effects}

Since H$\alpha$ is affected by dust extinction in the optical range, it is essential to consider dust effects on emission line flux measurements in order to derive accurate SFR results. As described in Sec. \ref{sec:dust_extinction}, the most robust way to estimate dust extinction in a galaxy is by measuring the flux ratio between H$\alpha$ and H$\beta$, also known as the Balmer decrement. One caveat of this method is that the H$\beta$ line is often very faint and undetectable, particularly for smaller, fainter galaxies such as the sources in our sample. To overcome the absence of H$\beta$ lines in our sample, the Balmer decrement for galaxies of similar properties was derived from the Sloan Digital Sky Survey (SDSS) \citep[]{york2000sloan} and used to estimate the typical Balmer decrement of star-forming galaxies in a mass bin. 
{This approach assumes that the median Balmer decrement trend with stellar mass also applies to the MUSE sample.}

{The total stellar mass of star-forming galaxies correlates with average metallicity, size and dust mass 
\citep{beeston18}. 
So naturally, the dust extinction is lower in low-mass galaxies because less sight lines go through high extinction regions.
Thus for our purposes, where we do not have individual Balmer decrement measurements, an average value as a function
of stellar mass is the best estimate we can make.}

\begin{figure}
    \centering
    \includegraphics[width=8.5cm]{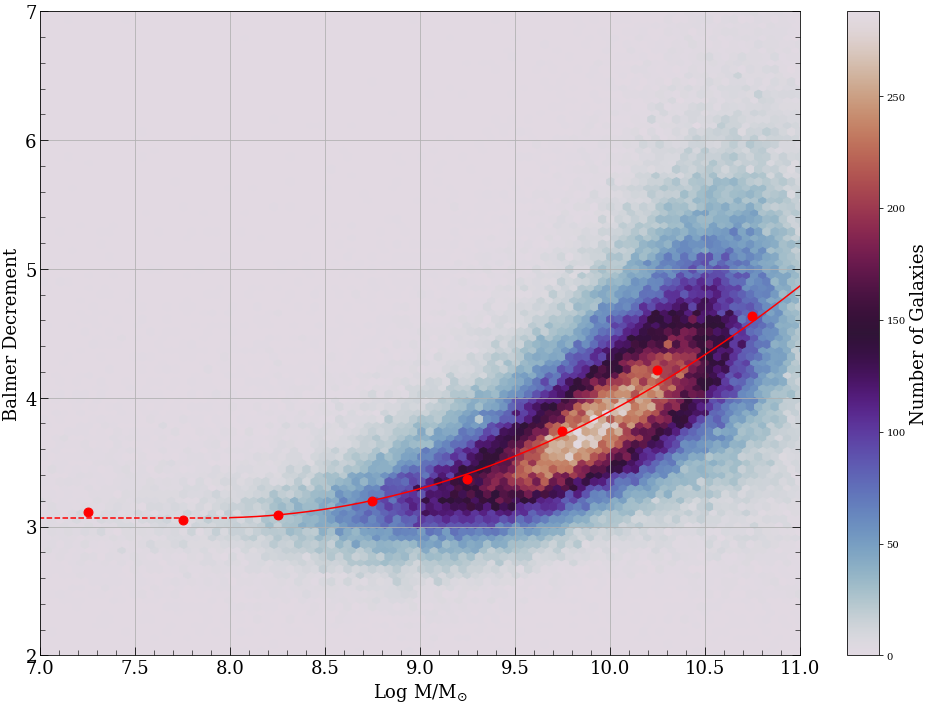}
    \caption{Balmer decrements derived using SDSS galaxies. The red solid line represents the median fit per each stellar mass bin of $\Delta \log M = 0.5$. The median fit was derived only for galaxies with stellar masses larger than $10^{8.5} \Msun$, as there is a non-negligible scatter at lower masses. The dotted line is the constant value used for bins with stellar mass lower than $10^{8} \Msun$.}
    \label{fig:balmer_fit}
\end{figure}

From SDSS, all the galaxies with measured H$\alpha$ and H$\beta$ emission line fluxes \citep{brinchmann2004physical,tremonti2004origin} and H$\alpha$ S/N $>$ 20 were considered. 
Then, the Balmer decrement for each galaxy was calculated, 
before determining the median value for each stellar mass bin of $\Delta \log$M = 0.5. 
From these points, a simple relation between Balmer decrement and stellar mass was derived 
{using a quadratic function for stellar masses larger than $10^{8.5} \Msun$, 
and assuming a constant decrement at lower masses. 
This provides a robust estimate that is sufficient for the scope of this study.}
The results are reported in Fig.~\ref{fig:balmer_fit}, where the red line shows the relation between Balmer decrement of star forming galaxies and their stellar mass, where it is possible to see how galaxies with lower stellar masses are significantly less affected by dust. The resulting values for Balmer decrement and dust extinction for each mass bin are reported in Table~\ref{tab:balmer}.

This dust-extinction-mass relation has been observed many times over the past decades, showing that the Balmer decrement, and thus dust extinction, is significantly larger in high mass galaxies \citep[]{dominguez2013dust, garn2010predicting}.
{For low-mass galaxies, the accuracy of the SFRD is not limited by the dust extinction estimate (only a correction factor of 1.2) whereas for high-mass galaxies ($>10^{10} \Msun$), it is arguably the limiting factor.}

\begin{table}
    \caption{Average Balmer decrements used in this study based on an analysis of SDSS star-forming galaxies, and the dust correction factor.}
    \centering
    \label{tab:balmer}
    \begin{tabular}{ccc}
        \toprule
        \textbf{Mass bin} & \textbf{Balmer Decrement} & \textbf{Dust correction factor}\\
        \midrule
         $<  7.5$  & 3.07 & 1.20\\
        7.5 -- 8.0 & 3.07 & 1.20\\
        8.0 -- 8.5 & 3.09 & 1.22\\
        8.5 -- 9.0 & 3.20 & 1.35\\
        9.0 -- 9.5 & 3.41 & 1.58\\
        \bottomrule
    \end{tabular}    
\end{table}

\subsection{H-alpha luminosity function}
\label{sec:lum_fun}
\begin{figure}
    \centering
    \includegraphics[width=8.5cm]{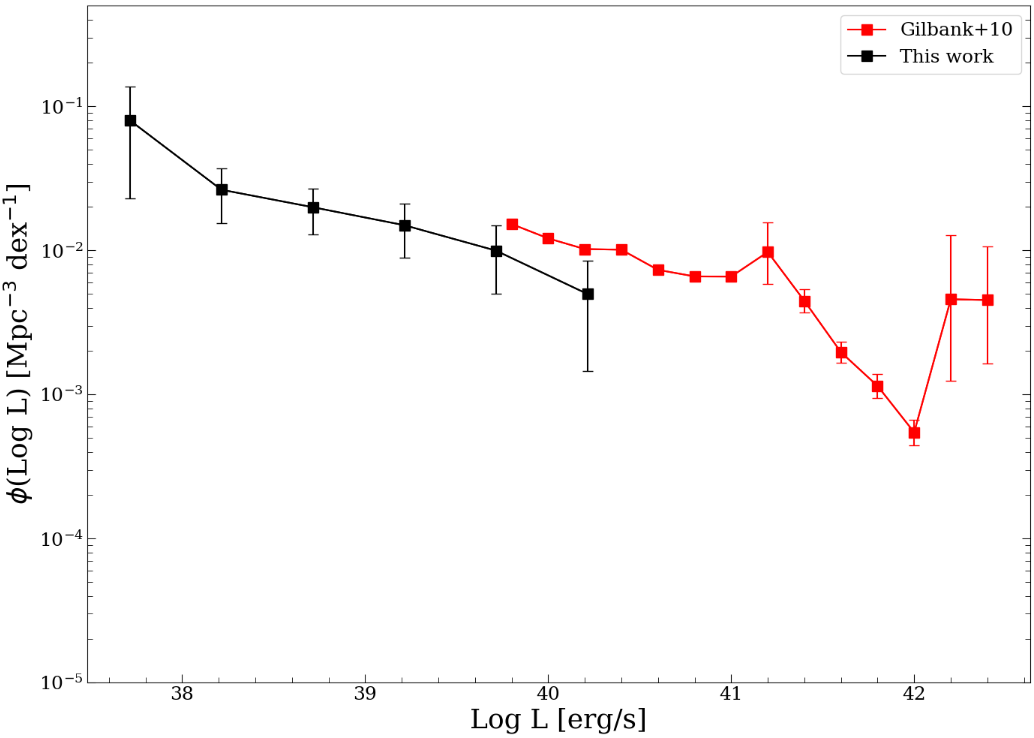}
    \caption{The observed H$\alpha$ luminosity function derived using the combined sample from MUSE Wide and MUSE HUDF compared to \citet{gilbank2010local}. This demonstrates how this work focuses on the fainter end of the galaxy population. The luminosities used in this plot are not dust corrected.}
    \label{fig:lum_func}
\end{figure}

In order to place the depth of the H-alpha selected sample into context, 
we compute the H$\alpha$ luminosity function using the $1/V_{max}$ values for the combined MUSE sample. 
The observed H$\alpha$ luminosity function (i.e.\ not corrected for dust) is shown in the plot in Fig.~\ref{fig:lum_func}, where it is compared with the results obtained by \cite{gilbank2010local}. 
This shows that the MUSE sample is probing to significantly fainter luminosities than the 
SDSS sample of \citeauthor{gilbank2010local}\ 
The latter sample was restricted to $0.032 < z < 0.2$. 
The SDSS spectra are reasonably well calibrated and the aperture corrections were made using the $u$-band in the higher-S/N SDSS Stripe 82 region. 
{The offset evident in the figure is likely due to the underdensity of the MUSE sample discussed in \S~\ref{sec:discussion}.}



For the GAMA sample, estimating H$\alpha$ luminosity functions is significantly harder because of the
smaller aperture used for the majority of spectra and the less accurate spectrophotometry 
(see \citealt{hopkins2013} for GAMA spectroscopic analysis). 
\citet{gunawardhana2013} compute dust-corrected H$\alpha$ luminosity functions for a similar GAMA sample and compare with other surveys as well. 
They discuss the problem of bivariate selection and the discrepancies between various results. 

\section{Results: the SFRD as a function of mass}
\label{sec:results}

\begin{figure}
    \centering
    \includegraphics[width=8.5cm]{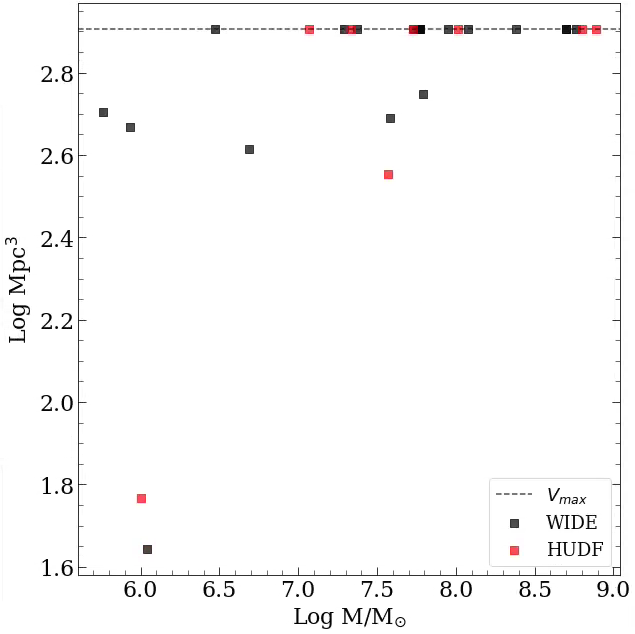}
    \caption{The maximum volume for each galaxy in the MUSE Wide and MUSE HUDF sample versus redshift. These are used for the estimation of the SFRD and luminosity function using the 1/V$_{max}$ method.}
    \label{fig:v_max}
\end{figure}

From the dust corrected H$\alpha$ line fluxes measured by MUSE Wide and MUSE HUDF, the SFRs of the galaxies were calculated using the calibration shown in Eq.~\ref{eq:SFRCalzetti}. Two star-forming galaxies, one from MUSE Wide and one from MUSE HUDF, although they have $OIII$ emission lines, do not present any confirmed H$\alpha$ emission, so they were excluded from the analysis, resulting in a total of 27 sources (19 from Wide and 8 from from HUDF). Before deriving the SFRD, the maximum volume of the survey was calculated using the detection limits determined previously. Fig. \ref{fig:v_max} shows the maximum volume of each galaxy, with $V_{max}$ derived considering the total volume of the two surveys combined, as described in sec \ref{sec:sfr_vmax}. The plot shows that the galaxies with the lowest stellar masses tend to have lower $V_{max}$, which is consistent with the expectation, as they tend to have lower H$\alpha$ flux and thus, they would not be observed at further distances. 

The SFRD for the combined MUSE Wide and MUSE HUDF sample was derived using the $V_{max}$ method with Eq. \ref{eq:sfrd} described in Sec. \ref{sec:sfr_vmax}. Given the limited number of galaxies in the sample, the mass bins were chosen to be relatively wide, with a value of $\Delta \log M/\Msun = 1$. This was necessary in order to have a reasonable estimate of the uncertainty using weighted-Poisson errors for each bin (\citealt{WeightedPoissonEvents}; the variance is the sum of weights squared {with the weights as per Eq.~\ref{eq:sfrd}}). The SFRD was calculated starting from two different mass bins, $\log M/\Msun = 5$ and  $\log M/\Msun = 5.5$, resulting in two overlapping functions for the same sample.  

The SFRD was also calculated using a sample from the GAMA survey, a large magnitude-limited spectroscopic survey. Compared to MUSE Wide and MUSE HUDF, GAMA is shallower (and missing low-surface-brightness galaxies) but with a much larger area, comprising a vaster number of galaxies and thus also providing a good indication of the number density at lower stellar masses. The GAMA catalogue used contained measurements of spectroscopic redshifts from the SpecCat DMU, stellar masses from StellarMasses DMU, and SFR measurements at different timescales derived through SED fitting with MagPhys in the MagPhys DMU. SED measurements of SFRs were used instead of deriving the SFR using measurements of H$\alpha$ emission line flux measurements, as GAMA fiber spectroscopy may overestimate their values for dwarf galaxies \citep{richards2016sami}. Since only young, massive stars with ages $<$ 20 Myr produce H$\alpha$ emissions \citep{davies2016gama}, the median SFR at timescales of 10$^7$ and 10$^8$ were used in this analysis (from the MagPhys table). After ensuring that only galaxies with valid measurements were included, the final sample from GAMA consisted of 7579 galaxies at redshift $z < 0.06$ from the equatorial regions that are highly complete (179 deg$^2$).

\begin{figure*}
    \centering
    \includegraphics[width=14cm]{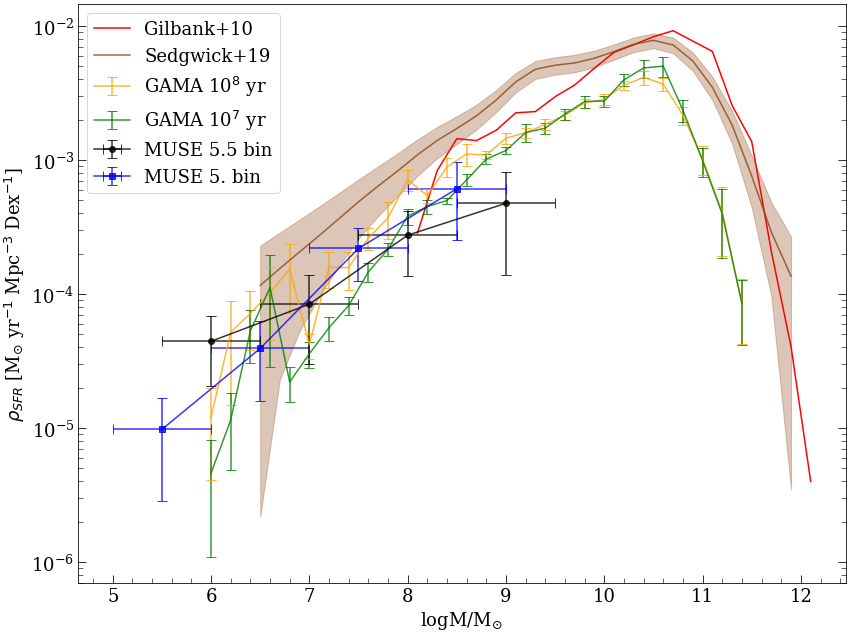}
    \caption{{Comparison between SFRD estimates from this and previous work: the combined MUSE sample (using two different binning schemes); the GAMA sample (using two different MagPhys timescales); the SDSS H$\alpha$ Balmer-decrement corrected measurement from \citet{gilbank2010local}; and the measurement using CCSNe from \citet{sedgwick2019galaxy}.
    The MUSE results use two different starting points, $\log M/\Msun = 5$ and  $\log M/\Msun = 5.5$, resulting in two overlapping functions.}}
    \label{fig:sfrd}
\end{figure*}

The results for the SFRD using the combined MUSE sample at different mass bins and using the GAMA sample at different timescales can be seen in Fig. \ref{fig:sfrd}. The plot shows that the samples appear to concur, with a steady decline at the lower mass end of the SFRD($M$). The decline may be quantified by assuming a linear regression between SFRD and stellar mass in log space at lower stellar masses, and measuring its slope $\gamma$, i.e. 
\begin{equation}
\label{eq:sfrd_slope}
    \frac{d\,\mathrm{SFRD}}{d \log M}  \propto M^\gamma \mbox{~~~.}
\end{equation}
The linear fit was calculated using the standard least-squares fitting method, where the diagonal elements of the covariance matrix correspond to the variances of the slope and intercept. In our analysis, the resulting values of the slope in SFRD are $\gamma$ = $0.57 \pm 0.16$ for the combined MUSE sample (note that for this value is derived using stellar mass bins of $\Delta \log M/\Msun = 0.5$), $\gamma$ = $0.73 \pm 0.14$  for the GAMA sample with median SFR at timescales of 10$^{7}$ yr, and $\gamma$ = $0.65 \pm 0.11$ for the GAMA sample median SFR at timescales of 10$^{8}$ yr. The SFRD for the GAMA sample were only fitted for $M < 10^8 \Msun$. 

The contribution of low-mass galaxies to the cosmic SFRD
has been measured in this paper. The slope of the SFRD is related to the slope of the GSMF, which is commonly modeled using a Schechter function so that the linear number density is proportional to M$^\alpha$, with $\alpha$ representing the slope of the GSMF at the faint end. Hence, the number density per logarithmic mass bin is given by
\begin{equation}
\label{eq:gsmf_slope}
    \frac{d\,N}{d \log M} \propto M^{\alpha+1} \mbox{~~~.}
\end{equation}

Considering the relation between SFR and stellar mass of star-forming galaxies, the SFR of a galaxy is approximately proportional to its stellar mass so that 
\begin{equation}
\mathrm{SFR} \propto M^\beta \mbox{~~~,}
\label{eq:beta}
\end{equation}
with $\beta \sim 1$, at masses below about $10^{10} \Msun$, along the star-forming main sequence \citep{brinchmann2004physical,lee2015,mcgaugh2017}. 
Since the SFRD is the number density times the mean SFR (e.g.\ \citealt{behroozi2013average}), 
the relationship between the slopes of the SFRD, SFR-mass relation and GSMF (of star-forming galaxies) can be expressed as: 
\begin{equation}
\label{eq:alpha_gamma}
    \gamma = \alpha + \beta + 1 \mbox{~~~.}
\end{equation} 
{This is accurate over any mass range where there is no significant break in the power laws.} 

{It is possible to estimate the GSMF directly from these samples, however, 
they are expected to be incomplete even for star-forming galaxies at low masses.
This especially true for H$\alpha$ selection which is highly biased toward high-SFR galaxies, with significantly reduced detectability for low SFRs
(lower luminosity, equivalent width and surface brightness).
Instead if we assume $\beta \simeq 1$ along the main sequence of star-forming galaxies, 
the above relationship (Eq.~\ref{eq:alpha_gamma}) becomes $\alpha \simeq \gamma - 2$.}
Therefore, the values of the slope of the SFRD imply $\alpha$ in the range $-1.5$ to $-1.2$. These values are consistent with previous studies of the GSMF \citep{moustakas13,wright2017galaxy,thorne2021,driver2022galaxy}, providing further evidence of no turnover at lower masses. 

\section{Discussion}
\label{sec:discussion}

The SFRD results from the combined sample using the MUSE Wide and MUSE HUDF catalogues, and the results obtained using the catalogues from the GAMA survey, show the behaviour of the SFRD down to stellar masses of $\thicksim  10^{5.5} M_\odot$ with MUSE and $\thicksim$ 10$^6$ M$_\odot$ with GAMA, pushing our understanding of the SFRD to the faintest galaxies. The results from this work indicate that {low-mass galaxies ($\sim 10^{6.5} \Msun$) contribute about 1\% of the cosmic SFRD (per dex) relative to massive galaxies at the peak of the SFRD($M$)}. 

The MUSE sample benefited from accurate H$\alpha$ emission line flux measurements, thanks to MUSE Wide and MUSE HUDF advanced IFU spectroscopy, therefore the SFR for this sample was derived using the calibration described in Sec. \ref{sec:ha_line}. However, since the sample only contained a limited number of 27 galaxies, the SFRD could only be derived using large stellar mass bins of $\Delta \log M/\Msun = 1$, in order to ensure a reasonable Poisson error. Nonetheless, the results show reasonable agreement with the extrapolation based on previous studies, and with the SFRD estimated using GAMA.

The SFRs used for the analysis of the GAMA sample were derived via SED fitting using MagPhys \citep[]{da2008simple}, as the H$\alpha$ emission line measurements for low-mass galaxies in GAMA might be inflated and too susceptible to corrections required for aperture effects. However, its large area allowed us to create a sample of 7579 star forming galaxies at $z < 0.06$. As shown in Fig. \ref{fig:sfrd}, the slope of the GAMA SFRD is similar to the one found using MUSE at low masses. 
There is an increase in the SFRD derived from GAMA as stellar mass increases up to $\thicksim 10^{10.5} M_\odot$. 
There is a sharp decline at the higher end of the function as expected from the exponential cutoff in the GSMF and from the higher fraction of quenched galaxies
at higher masses. 

Previous studies encountered difficulties in estimating the SFRD below $\thicksim 10^8 M_\odot$, as dwarf galaxies are challenging to observe. \cite{gilbank2010local} derived the SFRD down to stellar masses of $\thicksim 10^8 M_\odot$ using H$\alpha$, [OII], and $u$-band luminosities from the Sloan Digital Sky Survey (SDSS), whereas \cite{sedgwick2019galaxy} managed to estimate the SFRD at $\thicksim 10^7 M_\odot$ using CCSNe as "signposts" to low surface brightness galaxies to constrain their abundance and contribution to the total star formation rate density (SFRD) at lower masses. As shown in Fig. \ref{fig:sfrd}, both studies suggested a constant decline in SFRD, similarly to the results found in this work. 

\begin{figure*}
    \centering
    \includegraphics[width=14cm]{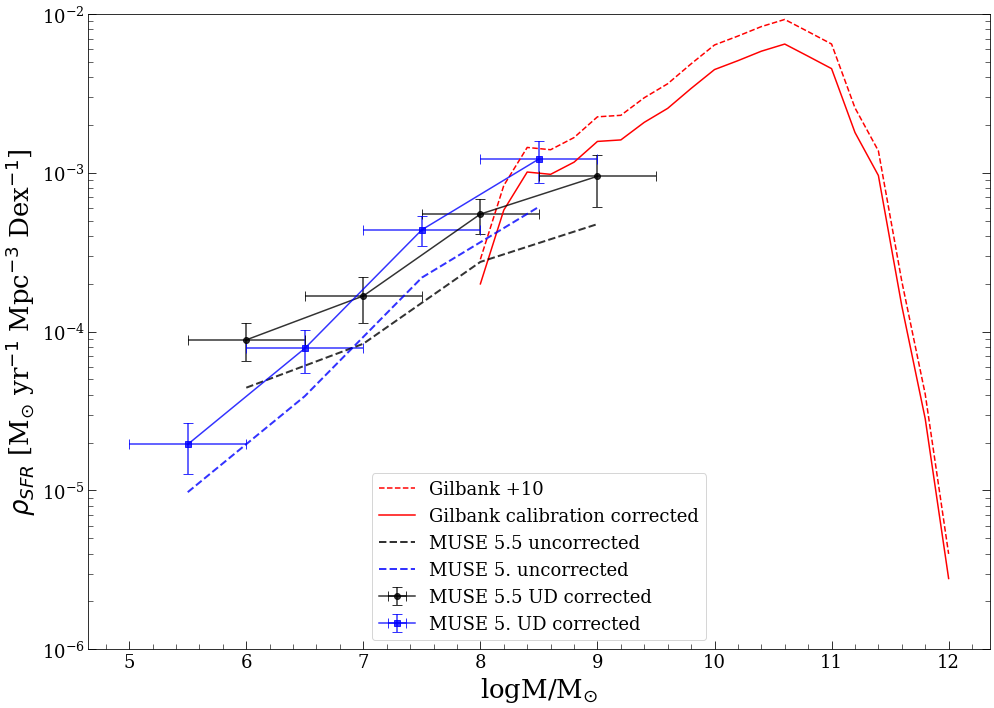}
    \caption{{SFRD with respect to stellar mass using H$\alpha$ line emission. 
    The MUSE combined sample is shown with and without a correction for the estimated underdensity (UD) of the sample.
    The uncorrected and UD corrected results are both shown for two different binning schemes starting at 5.0 and 5.5 in log mass. 
    The \citet{gilbank2010local} result is shown with and without a correction to convert their SFR-H$\alpha$ calibration 
    to that of Eq.~\ref{eq:SFRCalzetti}.}}
    \label{fig:sfrd_ud_corrected}
\end{figure*}

{The SFRD measurements of \citet{sedgwick2019galaxy} and \citet{gilbank2010local} do appear offset from
our work. However, \citeauthor{sedgwick2019galaxy} used a factor of two correction at low galaxy masses for 
the visibility of CCSNe, and there is uncertainty over classification of CCSNe for some of their sample. 
This also requires a CCSNe rate to SFRD calibration. So this SFRD from CCSNe may be overestimated. 
\citeauthor{gilbank2010local} used a different calibration of H$\alpha$ luminosity to SFR (from \citealt{kennicutt1998global}). 
This SFRD is shown adjusted to our calibration ($\times\,0.7$) in Fig.~\ref{fig:sfrd_ud_corrected}.} 

A constant decline in SFRD at lower stellar masses could have great significance in our understanding of the GSMF and galaxy evolution. 
In fact, the results ($\gamma < 1$) suggest that there is no turn-over at the faint end of the GSMF down to $\sim 10^{6} \Msun$. 
{This is based on the assumption that the main sequence slope (Eq.~\ref{eq:beta}) is approximately unity 
e.g.\ \citet{lee2015}.
Or in other words, the mean specific SFR of star-forming galaxies is similar for mass ranges below $10^{10} \Msun$
\citep{james2008halpha}. 
This is expected if these galaxies have formed their stars quasi-continuously since galaxy formation began. 
}

{
A reasonable estimate of the SFRD does not require that we find all star-forming galaxies in a given volume.
It is plausible that low-mass star-forming galaxies may go through lull periods (but not quenched)
which would reduce their H$\alpha$ luminosity significantly and thus not appear in H$\alpha$-selected samples. 
This is the duty cycle of star formation and it is likely more extreme at low masses. 
In effect, the SFRD over a cosmological volume is an estimate of the time average over a population of star-forming galaxies.
This is because we do not expect the chance of being in a lull or burst phase being correlated with other star-forming dwarf galaxies.}

The implications of our findings highlights the importance of low-mass galaxies in various cosmological processes. 
The slope of the SFRD suggests a steep faint-end slope for the GSMF, consistent with prior studies (e.g., \citealt{wright2017galaxy}). This highlights the importance of low-mass galaxies in the star formation budget and their potential role in early cosmic epochs. 

The volume of the MUSE sample is only 804 Mpc$^3$. Therefore this is potentially subject to significant cosmic variance. To test this, we consider previous observations of the GSMF from the GAMA survey \citep[]{baldry2012galaxy, wright2017galaxy}. Measurements of the number density for galaxies more massive than $10^8 \Msun$ are not affected by low-surface-brightness incompleteness in GAMA. Integrating the number density from the high-mass end, we obtain between 0.028 and 0.036 galaxies per Mpc$^3$ by changing the log-mass limit between 7.9 to 8.1. This allows for some uncertainty in how stellar mass is estimated. 

As the volume of the combined MUSE sample is 804 Mpc$^3$, the expected number of galaxies with stellar masses higher than $10^8 \Msun$ should be between 22 and 29. However, the total number of galaxies above this mass limit, including passive (non star-forming) galaxies detected using AUTOZ \citep[]{baldry2014galaxy}, in the volume of interest is only 12.  These results suggest that the volume is underdense. If we assume that the number density of galaxies below $10^8 M_\odot$ scales in the same way, we can account for the underdensity by multiplying the values of the SFRD by about two.

This simple correction can be seen in the plot reported in Fig. \ref{fig:sfrd_ud_corrected}, where it can be seen how by accounting for underdensity, the SFRD is more consistent with the values derived in previous studies \citep[]{sedgwick2019galaxy,gilbank2010local}. These results suggest that a larger sample is required to have a conclusive understanding of the low-mass galaxy population and their impact on the SFRD and GSMF. These results may also provide a useful benchmark for comparisons with future deep field surveys and cosmological simulations.

\section{Summary}
\label{Sec:summary}

This study investigates the contribution of star-forming galaxies to the SFRD as a function of stellar mass at $z < 0.2$, using MUSE Wide, MUSE HUDF, and GAMA surveys. By combining deep spectroscopic coverage and wide-area data, we probe the faint end of the SFRD down to stellar masses of $10^{5.5} \, \mathrm{M_\odot}$, providing new insights into the low-mass galaxy population. In summary, the results from this work show that:
\begin{itemize}
    \item The MUSE Wide and MUSE HUDF surveys have a total volume of 804 Mpc$^3$. Their exceptional depth allowed us to detect 27 star-forming galaxies with H$\alpha$ emission lines at redshifts $z < 0.2$. On the other hand, the sample from the GAMA survey benefited from a much larger number of galaxies, with a total of 7579 galaxies at redshift $z < 0.06$. 
    {For GAMA, we rely on the photometric estimates of the SFRs from SED fitting because the spectroscopic H$\alpha$ measurements require significant aperture corrections.}

    \item The H$\alpha$ luminosity function reported in Fig.~\ref{fig:lum_func} was derived using a combined sample of galaxies from MUSE Wide and MUSE HUDF down to values of $\thicksim$ 10$^{38}$ erg$/$s. {This shows that we push to significantly lower luminosities in H$\alpha$ compared to those used for the SFRD measurements of \cite{gilbank2010local}.} 
    
    \item The SFRD with respect to stellar mass distribution was estimated for the MUSE Wide and HUDF combined sample, using SFR H$\alpha$ measurements, and the GAMA sample, using SFRs values derived with MagPhys SED fitting. The results shown in Fig. \ref{fig:sfrd} suggest that there is a constant slope in SFRD at lower masses with values of the slope $\gamma$ (Eq.~\ref{eq:sfrd_slope}) between 0.57 to 0.73. 
    {This quantifies the contribution of low-mass galaxies to the cosmic SFRD.}
    
    \item The slope in SFRD($M$) implies that the faint-end slope in the GSMF, $\alpha$, is in the range $-1.5$ to $-1.2$ (see Eqs.~\ref{eq:gsmf_slope}--\ref{eq:alpha_gamma} for this argument), consistent with values found in GAMA \citep[]{baldry2012galaxy,wright2017galaxy}, providing important evidence of the absence of a turn-over at the faint end of the GSMF. 
    
    \item When considering the number density of galaxies more massive than $10^8 M_\odot$ using the GAMA GSMFs compared with the number of galaxies detected at $z<0.2$ in MUSE Wide and MUSE HUDF, 
    there is a significant discrepancy between the number densities. 
    This implies that the volume considered in MUSE is underdense, with about a factor-of-two fewer galaxies than the cosmic mean.
    
    \item The results could be improved with larger surveys, allowing us to more accurately investigate the faint end of the SFRD and GSMF, and with comparisons with cosmological simulations. For example, the Rubin observatory can be used to detect and characterize the SFRD using CCSNe; while Euclid imaging (to detect low surface-brightness galaxies) plus redshifts from 4MOST provide a deeper galaxy redshift survey. And we could also use more MUSE fields (blind to low-redshift volume) for H$\alpha$ selected galaxy samples. 
\end{itemize}

\section*{Acknowledgements}

We thank the anonymous referee for thoughtful comments that significantly improved the paper. 
We would like to express our sincere gratitude to Prof.\ Phil James for his invaluable guidance and support throughout this work. 
We also thank the MUSE collaboration for providing the data used in this research, particularly from the MUSE WIDE and MUSE HUDF surveys. 
Finally, GM acknowledges the financial support from the Science and Technology Facilities Council (STFC) for funding his PhD.

\section*{Data Availability}

The data underlying this article are all publicly available. The MUSE WIDE survey data can be accessed from \url{https://musewide.aip.de/}, and the MUSE HUDF DR2 survey data are available at \url{https://amused.univ-lyon1.fr/}. Additionally, the data from the GAMA survey are accessible from \url{https://www.gama-survey.org/dr4/}.



\bibliographystyle{mnras}
\bibliography{mnras_template} %







\bsp	
\label{lastpage}
\end{document}